\documentclass[prl,nobibnotes,twocolumn,showpacs]{revtex4}
\usepackage{graphics,color,array,dcolumn}
\usepackage{calc}
\usepackage[T1]{fontenc}
\usepackage[latin1]{inputenc}
\usepackage{amsmath}
\usepackage{amssymb}
\usepackage{xspace}
\allowdisplaybreaks[1]

\newcommand{\be}{\begin{equation}}
\newcommand{\ee}{\end{equation}}
\def\bea{\begin{eqnarray}}
\def\eea{\end{eqnarray}}

\makeatother

\begin{document}
\title{Are there scaling solutions in the $O(N)$-models for large $N$ in $d>4$?}
\author{Roberto Percacci}
\email{percacci@sissa.it}
\affiliation{SISSA, via Beirut 4, I-34014 Trieste, Italy, and INFN, 34127 Sezione di Trieste, Italy}
\author{Gian Paolo Vacca}
\email{vacca@bo.infn.it}
\affiliation{INFN, Sezione di Bologna, Via Irnerio 46, I-40126 Bologna, Italy}
\begin{abstract}
There have been some speculations about the existence of critical unitary $O(N)$-invariant scalar field theories in dimensions $4<d<6$ and for large $N$. 
Using the functional renormalization group equation,
we show that in the lowest order of the derivative expansion,
and assuming that the anomalous dimension vanishes for large $N$,
the corresponding critical potentials are either unbounded from below
or singular for some finite value of the field.
\end{abstract} 
\maketitle
\section{INTRODUCTION}
It has been suggested recently that the linear $O(N)$ scalar models 
may be conformal and unitary for sufficiently large $N$ in dimensions $4<d<6$ \cite{fgk}.
The argument is based on a model with a potential that
is unbounded from below and the conjectured fixed point has a field 
anomalous dimension of order $1/N$.
The presence of a non trivial UV fixed point would imply an unexpected
asymptotically safe behavior of such theories.
Here we consider the $O(N)$ models from the point of view of the functional renormalization group,
which has proved very successful in analyzing critical phenomena in dimensions $d<4$.
We find by means of analytic and numerical methods
that within certain approximations described below, for large $N$,
scaling solutions (fixed points of the action functional)
are either unbounded from below or singular and not globally defined.

The tool we use is the 1-PI generating functional, or Effective Average Action (EAA) $\Gamma_k$, which is defined as
the ordinary effective action but with an additional term in the action (infrared cutoff)
suppressing the contribution of modes with momentum lower than $k$.
It is convenient to choose a quadratic cutoff operator, whose kernel is usually denoted $R_k$,
so that the exact flow equation for the EAA in the RG time $t=\log(k/k_0)$ is given by
$\partial_t \Gamma_k = \frac{1}{2} {\rm Tr}[( \Gamma_k^{(2)}+R_k)^{-1} \partial_t R_k]$
\cite{wett1,morris,ber}.
This equation can be taken as an alternative definition of a QFT:
given that $R_k\to0$ for $k\to0$, the solution of the equation
starting from a given bare action at a UV scale,
gives the usual quantum effective action at $k=0$.

For applications to critical phenomena one expects an expansion in derivatives
to be a good approximation.
The most general form of the (Euclidean) EAA containing up to two derivatives is
\be
\int d^dx\left[\frac{Z_k(\rho)}{2}\partial_\mu\phi^a\partial^\mu\phi^a
+\frac{Y_k(\rho)}{4}\partial_\mu\rho\partial^\mu\rho
+V_k(\rho)\right]\ ,
\ee
where $Z_k$, $Y_k$ and $V_k$ are functions of $\rho=\phi^a\phi^a/2$ depending
on the external cutoff scale $k$.
The zeroth order of the derivative expansion,
called the Local Potential Approximation (LPA) consists in setting $Z_k=1$ 
(and the anomalous dimension $\eta=0$), $Y_k=0$
and studying only the running of $V_k$.
The first order consists in keeping the running of all three functions,
treating the anomalous dimension $\eta$ as a free parameter.
In $d=3$ this has been studied, with a power-law cutoff, in \cite{morris3}.
From these earlier studies we know that the
contribution of $Y_k$ to the running of $V_k$ and $Z_k$ is of order $1/N$, 
so that it can be neglected in the large $N$ limit.
Moreover, in the same limit, the only known solutions of the fixed point equations for 
$V_k$ and $Z_k$ have a field--independent $Z_k$ with anomalous dimension $\eta=0$. 
These clearly should correspond to the large-$N$ limit of an $N$-dependent family of solution with $\eta$ suppressed for large $N$.
No other solutions, with a non vanishing $\eta$ at $N=\infty$, are known up to now.
The next order, where one keeps terms with four derivatives,
has been studied only in the case $N=1$ \cite{canet,LZ}.

In most other works the functions $Z_k$ and $Y_k$ are expanded around a constant
field configuration $\rho_0$ , typically at the minimum of the potential, and only the first term is retained.
The wave function renormalization constants of the Goldstone and radial modes
are defined by $Z_k=Z_k(\rho_0)$ and $\tilde Z_k=Z_k(\rho_0)+\rho_0 Y_k(\rho_0)$
and the anomalous dimensions are defined as
$\eta=-\partial_t Z_k/Z_k$ and $\tilde\eta=-\partial_t\tilde Z_k/\tilde Z_k$ \cite{wett2,tw,tl,btw}.
This setup, which is sometimes called LPA', makes contact with
the perturbative formulae for the anomalous dimension and extends them
somewhat in a non-perturbative way.
In this work we confine ourselves mostly to the LPA.

\section{CALCULATIONS}

In order to have a beta functional for $V_k$ in closed form we shall
consider the piecewise linear cutoff discussed in \cite{optimized},
which allows the loop integrals to be performed exactly.
Defining the dimensionless field $\tilde\phi^a=k^{(1-d/2-\eta/2)}\phi^a$, the dimensionless
potential $\tilde V(\tilde\phi)=k^{-d}V$ and rescaling 
$\rho=N c_d\tilde\phi^2/2$, $u=N c_d\tilde V$, with $c_d=(1-\eta/(d+2))(4\pi)^{-d/2}/\Gamma(d/2+1)$, 
the flow equation for the potential assumes the simple form
\be
\label{udot}
\dot u=-d u(\rho)+(d-2+\eta)\rho u'(\rho)+\frac{1}{1+u'(\rho)}+A(N)\ ,
\ee
where
\be
\label{udotA}
A(N)=\frac{1}{N}\left(\frac{1}{1+u'(\rho)+2\rho u''(\rho)}-\frac{1}{1+u'(\rho)}\right)\ .
\ee
The dot and prime denote partial derivatives with respect to $t$ and $\rho$ respectively.

In order to first establish a link with the work of \cite{fgk} we can make
a quartic ansatz for the potential $u=\lambda_2\rho+\lambda_4\rho^2$
and insert it in (\ref{udot}) to obtain beta functions for $\lambda_2$ and $\lambda_4$.
These turn out to have a fixed point at 
\bea
\small
\lambda_{2*}&=&\frac{-(d-4)(N+2)}{(d-8)N+2d-40}\ ;
\nonumber\\
\lambda_{4*}&=&\frac{16(d-4)N(N+8)^2}{((d-8)N+2d-40)^3}\ .
\label{polysol}
\eea
Taking into account the rescalings, $\lambda_{4*}$ agrees with Eq.(2.7) of \cite{fgk}
to first order in $\epsilon=d-4$.
An approximate expression for the anomalous dimension is obtained from the flow of $\Gamma^{(2)}$:
one has $\eta=c_d\frac{4\bar\rho u''(\bar\rho)^2}{(1+2\bar\rho u''(\bar\rho))^2}$
where $\bar\rho=-\lambda_2/(2\lambda_4)$ is the stationary point of the potential.
Using the fixed point couplings,
\be
\eta=\frac{128(d-4)^2(N+2)(N+8)^2}{((d-8)N+2d-40)^2((3d-16)N+6d-56)^2}
\ee
This is twice the anomalous dimension of the fields $\phi^a$,
and agrees with Eq. (2.4) of \cite{fgk} to first order in $\epsilon$.
One notices that $\eta$ is of order $1/N$.
From this and partial results of the second order derivative expansion quoted above, 
we shall assume that the anomalous dimension is negligible in the large $N$ limit.
We discuss this assumption further in the concluding remarks.

Our strategy is to solve Eq.~(\ref{udot}) without expanding the potential.
The fixed point equation $\dot u=0$ is a second order differential equation 
with a fixed singularity in $\rho=0$, so that analytic solution are parametrized by 
a single integration constant $\sigma=u'(0)$.
Around the origin analytic solutions can be written as
$
u(\rho)=\frac{1}{d (1+\sigma)}+ \sigma \rho 
+O\left(\rho^2\right).
$
Such an expansion is valid only within a certain radius of convergence.
One can integrate the equation numerically with this 
boundary condition at a very small $\rho$.
For any fixed $\sigma$, integration fails at some maximum value $\rho_{max}(\sigma)$.
In $d=3$ this function has a sharp spike at a negative value of $\sigma$ that corresponds to
the Wilson-Fisher fixed point, and althought it is in practice not possible
to continue this solution numerically to infinity, it can be matched smoothly
to a solution satisfying the right asymptotic behavior at large $\rho$~\cite{morris3}.

\begin{figure}[htp]
\centering
\resizebox{0.8\columnwidth}{!}{
\includegraphics{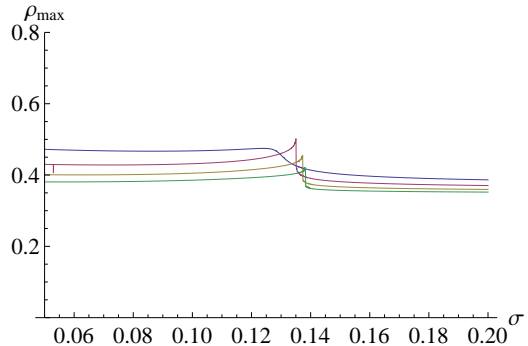}}
\caption{Top to bottom: the function $\rho_{max}(\sigma)$ for $N=2000$, $4000$, $8000$, $16000$.}
\end{figure}

One can repeat this analysis for any $d$ and in particular in $d=5$.
For $N\approx10^3$ the function $\rho_{max}(\sigma)$ has a smooth bump at $\rho\approx0.12$.
The bump becomes more pronounced as $N$ increases, and it moves to slightly
larger $\rho$. This behavior is shown in Fig.~1.
One could suspect that the peak corresponds to a scaling solution as $N\to\infty$.
Unfortunately, when $N\approx 10^5$ it becomes increasingly difficult to study numerically
what happens near the peak.

However the case $N=\infty$ can be treated analytically.
It is convenient to derive the fixed point equation once and to define $u'(\rho)=w(\rho)$ to obtain
\be
\label{eqlargeN}
-2 w(\rho)+(d-2)\rho w'(\rho)-\frac{w'(\rho)}{(1+w(\rho))^2}=0\,.
\ee
From this equation one immediately sees that apart from the trivial solution ($w=w'=0$)
the potential can have a stationary point only at $\rho_0=1/(d-2)$.
Deriving repeatedly Eq.(\ref{eqlargeN}) one determines the Taylor coefficients 
of the expansion of the solution around this point:
\be
w(\rho)=-\frac{d\!-\!4}{2} (\rho-\rho_0)+\frac{3}{8}\frac{(d\!-\!4)^3}{d\!-\!6}(\rho-\rho_0)^2+O\left((\rho\!-\!\rho_0)^3\right)
\nonumber
\ee
The general solution is given by the implicit relation
\be
\!\rho=C' w^{\frac{d}{2}-1}
\!+\frac{1}{(d\!+\!2)(1\!+\!w)^2}\!\!\phantom{a}_2F_1\left(\!\!1,2,2\!+\!\frac{d}{2},\frac{1}{1\!+\!w}\!\right).
\label{sol1}
\ee
For $d\not=2n$ the solution of Eq.~\eqref{eqlargeN} is most conveniently written in the form
\cite{marchais}
($C'=C-\frac{d \pi}{4}/\sin(d \pi/2)$).
\be
\rho(w)=C w^{\frac{d}{2}-1}
+\frac{1}{d-2}\!\!\phantom{a}_2F_1\left(2,1-\frac{d}{2},2-\frac{d}{2},-w\right)
\label{sol2}
\ee

\begin{figure}[htp]
\centering
\resizebox{0.7\columnwidth}{!}{
\includegraphics{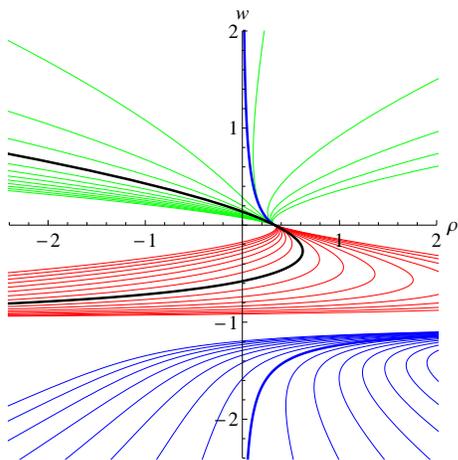}}
\caption{The solutions of Eq.(\ref{eqlargeN}) with $C$ real ($w>0$, green), $C$ imaginary ($-1<w<0$, red), $C'$ imaginary ($w<-1$, blue),
and the special solutions $C=0$ ($w>-1$, thick black) and $C'=0$ ($w>0$, thick blue).}
\end{figure}

The only real solution extending continuously through $w=0$ is the one with $C=0$,
which has the Taylor expansion given above. 
For $d=5$ the solution is the thick black curve in Fig.~2, enlarged in Fig.~3.
It intersects the $\rho=0$ axis at $w(0)\approx 0.1392$,
which corresponds to the accumulation point of the
peaks of Fig.~1, and at $w(0)\approx -0.5776$.
We see that $u''$ has a singularity at $\rho_{max}\approx 0.621$,
where the implicit solution cannot be inverted,
and the potential does not exist beyond this value.

The solutions corresponding to different values of the integration constant are also shown in Fig.~2.
The ones with real $C\not=0$ exist only for positive $w$ (green curves),
those for imaginary $C\not=0$ exist only in the range $-1<w<0$ (red curves)
and those of the form given in Eq.~\eqref{sol1} with $C'$ purely imaginary exist only for $w<-1$ (blue curves).
If ${\rm Im}(C')<0$ the latter form a continuum of global solutions covering the whole range
$-\infty<\rho<\infty$, therefore including the whole physical region.
However, these solutions have $u'<0$ everywhere so that the potentials $u$ are unbounded from below
and therefore are physically unacceptable.
The green and red curves also have 
$w'(\rho_0)=-(d-4)/2$ for $d>4$,
but some higher derivative diverges at this point,
for example in $d=5$, $u'''(\rho_0)=w''(\rho_0)$ is singular.
In conclusion, we have shown analytically
that there is no acceptable scaling solution in the LPA at $N=\infty$.

\begin{figure}[htp]
\centering
\resizebox{0.7\columnwidth}{!}{
\includegraphics{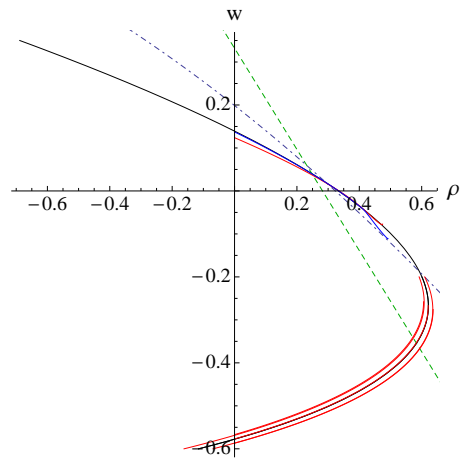}}
\caption{The exact solution for $N=\infty$ with $C=0$ (black curve) with superimposed (top to bottom):
the polynomial solution for N=2000 at order $\rho^2$ of Eq.(\ref{polysol}) (green, dashed) and $\rho^3$ (blue, dot-dashed);
numerical solutions for 
$N=8000$ (with $\sigma=0.137$, blue curve) 
$N=2000$ (with $\sigma=0.124$, red curve),
corresponding to the "peaks" in Fig.~1;
numerical solutions for $N=2000$ and $\sigma=-0.567,-0.577,-0.587$ (red, continuous).
}
\end{figure}

For large but finite $N$ we have studied
the solutions numerically integrating the flow equation
starting either from $\rho=0$ or from large $\rho$.
The solutions corresponding to the peaks in Fig.1 are very close to
the exact solution with $C=0$ at infinite $N$ 
up to some maximal value for $\rho<\rho_{max}$, see Fig.~3.
The numerical evolution then starts to deviate and stops where $u''$ diverges.
For $N\approx 8000$ the curves are nearly indistinguishable.
Similarly one can analyse the numerical evolution from initial conditions
near $w\approx -0.577$. Nothing special happens here: 
the function $\rho_{max}(\sigma)$ shows no bumps or peaks for negative $-1<\sigma<0$.
The evolution for varying initial conditions produces curves that are close to the
large-$N$ solution, all terminating in a singularity of $u''$, see Fig.~3.

For large $\rho$ the solutions of Eq.(\ref{udot}) have an asymptotic expansion
$u =A \rho^\alpha+\frac{1}{d A(2\alpha-1)}\left(1-\frac{1}{N} \frac{2(\alpha-1)}{2\alpha-1}\right)\rho^{1-\alpha}+O\left(\rho^{2-2\alpha}\right)$
where $\alpha=d/(d\!-\!2\!+\!\eta)$ and $A$ is an arbitrary constant.
If a global solution existed, 
one should be able to find it from a numerical integration with this asymptotic behavior. 

As in the exact large-$N$ case,
we find it more convenient to analyze this problem in terms of $\rho(w)$ which satisfies
the fixed point differential equation
\be
(d-2+\eta)\rho -(2-\eta) w \rho'-\frac{1\!-\!\frac{1}{N}}{(1\!+\!w)^2}
+\frac{1}{N}\frac{2\rho\,\rho''-3\rho'{}^2}{(2\rho\!+\!(1\!+\!w)\rho')^2}=0
\nonumber
\ee
which determines the asymptotic behavior:
\be
\label{asbe}
\rho=A w^\beta +\frac{1-\frac{2}{N(2+\beta)}}{d+2-\eta} w^{-2}+O\left( w^{-2} \right)
\ee
where $\beta=(d-2+\eta)/(2-\eta)$.
In the large-$N$ limit this series correspond exactly to the one given by the expansion of expressions in Eqs.~\eqref{sol1} and \eqref{sol2}.
Solving numerically with $\eta=0$
we find that the function $\rho(w)$ never becomes zero.
Instead, it turns around and crosses $w=0$ without singularities near $w=1/(d-2)$,
then tends to $+\infty$ at $w=-1$.
Inverting this relation, the solutions for $w(\rho)$ become
singular before reaching $\rho=0$, so they do not correspond to globally defined scaling solutions
(see Fig.~4).
These numerical studies show that the conclusion of the $N=\infty$
case extends also to large but finite $N$.

\begin{figure}[htp]
\centering
\resizebox{0.7\columnwidth}{!}{
\includegraphics{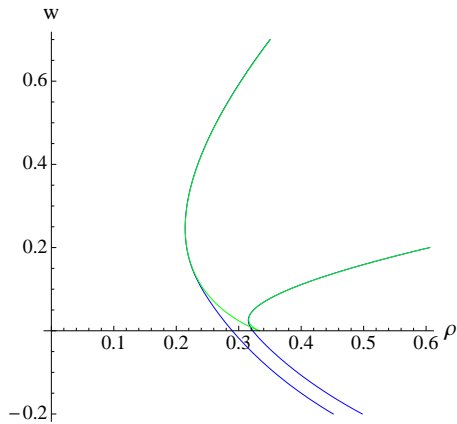}}
\caption{Two exact solution for $N=\infty$ ($w>0$, green curves) and the corresponding numerical solution 
matching the asymptotic behavior Eq.(\ref{asbe}) for $N=8000$ (extending to negative $w$, blue curves).}
\end{figure}

We worked throughout with a specific choice of the cutoff kernel $R_k$
and one may wonder whether different cutoff shapes could lead to qualitatively different results.
We have checked that similar results are obtained with a sharp cutoff.
For an arbitrary cutoff the fraction in the third term of Eq.~(\ref{udot}) can be approximated by
$C_1-C_2u'(\rho)$ for small $u'(\rho)$ and by $C_3/u'(\rho)$ for large $u'(\rho)$,
where $C_1$, $C_2$, $C_3$ are cutoff--dependent constants of order one
(the terms in $A(N)$ also behave in a similar way).
One cannot write the solutions in closed form for general cutoff, but for large and small
$\rho$ they have the same general behavior as the solutions discussed above,
so we expect our results to hold for any cutoff shape.
We have also repeated the analysis in other dimensions $d>4$ with similar results.

Finally, as a preliminary step towards a more complete analysis,
we have studied the solutions of Eq.(\ref{udot}) for $N=1000$
with nonvanishing anomalous dimension $\eta$ ranging between $-0.05$ and $0.05$.
The dependence of the solutions on $\eta$ is smooth and one always finds
the same qualitative behavior as with $\eta=0$.

\smallskip

\section{DISCUSSION} 
While the polynomial expansion of the potential is a useful tool,
it may sometimes be misleading.
Truncating the potential to a finite polynomial generally leads to spurious fixed points
and one has to carefully sift for the physically significant ones~\cite{morris-last}.
In the present case, Taylor expanding the potential
at second order in $\rho$ leads to (\ref{polysol}),
which corresponds to the dashed green curve in Fig.3.
Polynomials of higher order yield, among others, solutions that
become arbitrarily good approximations of
the exact solution $N=\infty$, $C=0$ (black curve in Fig.2 and 3).
These polynomial approximations exist for all $\rho$ but we have seen
that the real solution that they approximate does not.

If one keeps the whole Taylor series, and a recursive method is used to find 
the solution, one has to check the radius of convergence of the series.
For example, if one considers a general scalar potential and
writes the beta functions for the coefficients of its Taylor expansion
around the origin, then given an arbitrary value to $\sigma=\lambda_2$,
the beta functions can be solved iteratively yielding a fixed point potential \cite{hh}.
The existence of such solutions in $d=4$ would go against the expectation that linear scalar 
theory is trivial in dimension $d\geq4$.
Indeed it was shown by Morris \cite{morris1} that in $d=4$ these fixed point potentials 
become singular at some finite value of the field. 
On the other hand in $2<d<4$, only for a discrete set of values of $\sigma$ 
does the solution extend to infinity.
These global scaling solutions correspond to multicritical models 
(see \cite{morris2,codello} for the $Z_2$-invariant case,
\cite{cood} for some results in the general $O(N)$ case).

The functional method we have used here avoids expanding the solution.
We have used the minimal truncation that yields the desired information,
keeping the running of the potential.
In the large $N$ limit we have shown analytically that no global solution 
exists that is bounded from below. For the class of theories with $\eta=0$ at $N=\infty$ this result goes beyond the LPA.
For large but finite $N$ we have reached the same conclusion by studying
numerical solutions starting from $\rho=0$ or $\rho=+\infty$.
In both cases either $u''$ becomes singular at some finite $\rho$ 
or the potential has everywhere $u'<0$ and is therefore unbounded from below.
Note that this is also the case for the QFT model analysed in~\cite{fgk}.
Even though we concentrated on large $N$, 
we have found no evidence of a qualitative change of behavior as $N$ decreases.
This is in agreement with the analysis of \cite{nakayama},
which was based on the conformal bootstrap.
We have focused primarily on the case $d=5$ but a least in the LPA there is evidence that the conclusions hold for all $d\ge4$.

For a complete analysis at first order of the derivative expansion 
one would have to study the flow of the three functions $V_k$, $Z_k$, $Y_k$ as in \cite{morris3}.
The LPA' formulas indicate that $\eta$ is of order $1/N$,
making the existence of solutions with $\eta\not=0$ for $N=\infty$ seem unlikely.
In any case, if such a solution existed it would be in another
universality class from the one conjectured in \cite{fgk}.
For finite $N$ the system of equations for $V_k$, $Z_k$, $Y_k$ is completely coupled
and one cannot rule out the existence of solutions, but this we also consider to be unlikely,
since in lower dimensions all the qualitative features of the fixed point 
can already be seen in the LPA.
A possible exception may occur if, in presence of non global but singular solutions, the function $Z_k(\rho)$ has a singularity at the
same location as the potential. Then it may be legitimate to redefine the quantum field
in such a way as to shift the singularity to infinity.
We leave this to future investigations.

{\it Acknowledgements}. We would like to thank A. Codello, S. Giombi, I. Klebanov and D. Litim 
for discussions.
\goodbreak
\medskip


\end{document}